\documentclass[12pt]{article}
\usepackage[super]{natbib}
\usepackage{graphicx}

\newcommand{\rhessi}{\emph{RHESSI}}

\renewcommand{\deg}{\ensuremath{^\circ}}

\newcommand{\asec}{\ensuremath{''}}

\begin{document}

\title{Polarization of the prompt $\gamma$-ray emission from the
$\gamma$-ray burst of 6 December 2002}

\author{Wayne Coburn$^{*}$ \& Steven E. Boggs$^{*,\dag}$}

\date{Nature, 2003 May 22, {\bf 423}, 415-417}

\maketitle

{\small\noindent$^{*}$Space Sciences Laboratory, and\\
$^{\dag}$Department of Phyisics, University of California at Berkeley,\\
Berkeley, CA, 94720, USA}

\vspace*{0.5in}





{\bf Observations of the afterglows of $\gamma$-ray bursts (GRBs) have
revealed that they lie at cosmological distances, and so correspond to
the release of an enormous amount of energy\citep{mes02,par00}. The
nature of the central engine that powers these events and the prompt
$\gamma$-ray emission mechanism itself remain enigmatic because, once
a relativistic fireball is created, the physics of the afterglow is
insensitive to the nature of the progenitor. Here we report the
discovery of linear polarization in the prompt $\gamma$-ray emission
from GRB021206, which indicates that it is synchrotron emission from
relativistic electrons in a strong magnetic field.  The polarization
is at the theoretical maximum, which requires a uniform, large-scale
magnetic field over the $\gamma$-ray emission region. A large-scale
magnetic field constrains possible progenitors to those either having
or producing organized fields. We suggest that the large magnetic
energy densities in the progenitor environment (comparable to the
kinetic energy densities of the fireball), combined with the
large-scale structure of the field, indicate that magnetic fields
drive the GRB explosion.}

We used the \emph{Reuven Ramaty High Energy Solar Spectroscopic
Imager} (\rhessi)\citep{lin03} to make these$\gamma$-ray observations
of GRB021206.  \rhessi\ has an array of nine large-volume
(300\,cm$^3$) coaxial germanium detectors with high spectral
resolution, designed to study solar X-ray and$\gamma$-ray emission
(3\,keV--17\,MeV). \rhessi\ has high angular resolution (2\asec) in
the $\sim$1\deg\ field of view of its optics; however, the focal plane
detectors are unshielded, open to the whole sky. Thus, while the
chances are small that \rhessi\ will see a GRB in its imaging field of
view, it measures them frequently in the focal plane detectors
themselves. These observations provide highresolution spectra,
individual photon times and energies, as well as the potential for
polarization measurements. \rhessi\ is not optimized to act as a
$\gamma$-ray polarimeter, but several aspects of its design make it
the most sensitive instrument to date for measuring astrophysical
$\gamma$-ray polarization.

In the soft $\gamma$-ray range of $\sim$0.15--2.0\,MeV, the dominant
photon interaction in the \rhessi\ detectors is Compton scattering. A
small fraction of incident photons will undergo a single scatter in
one detector before being scattered and/or photoabsorbed in a second
separate detector, which are the events sensitive to the incident
$\gamma$-ray polarization. Linearly polarized $\gamma$-rays
preferentially scatter in directions perpendicular to their
polarization vector. In \rhessi\, this scattering property can be used
to measure the intrinsic polarization of astrophysical sources. The
sensitivity of an instrument to polarization is determined by its
effective area to scatter events, and the average value of the
polarimetric modulation factor\citep{nov75,lei97}, $\mu(\theta,E)$,
which is the maximum variation in azimuthal scattering probability for
polarized photons. This factor is given by $\mu = (d\sigma_{\perp} -
d\sigma_{\parallel})/(d\sigma_{\perp} + d\sigma_{\parallel})$,
where $d\sigma_{\perp}$,$d\sigma_{\parallel}$ are the Klein-Nishina
differential cross-sections for Compton scattering perpendicular and
parallel to the polarization direction, respectively, which is a
function of the incident photon energy $E_{\gamma}$, and the Compton
scatter angle $\theta$ between the incident photon direction and the
scattered photon direction. For a source of count rate $S$ and
fractional polarization $\Pi_{s}$, the expected azimuthal scatter
angle distribution is $dS/d\phi =
(S/2\pi)[1-\mu_{m}\Pi_{s}\cos(2(\phi-\eta))]$, where $\phi$ is the
azimuthal scatter angle, $\eta$ is the direction of the polarization
vector, and $\mu_{m}$ is the average value of the polarimetric
modulation factor for the instrument. Although \rhessi\ has a small
effective area ($\sim$20\,cm$^2$) for events that scatter between
detectors, it has a relatively large modulation factor in the
0.15--2.0\,MeV range, $\mu_{m}\approx0.2$, as determined by Monte
Carlo simulations described below.

In comparison with other $\gamma$-ray instruments (COMPTEL, BATSE)
that have attempted to measure polarization in the
past\citep{lei97,mcc96}, \rhessi\ has the major advantage of quickly
rotating around its focal axis (centred on the Sun) with a 4\,s
period. Rotation averages out the effects of asymmetries in the
detectors and passive materials that could be mistaken for a
modulation. Because polarimetric modulations repeat every 180\deg, any
source lasting more than half a rotation (2\,s) will be relatively
insensitive to the systematic uncertainties that typically plague
polarization measurements.  Finally, although the \rhessi\ detectors
have no positioning information themselves, they are relatively
loosely grouped on the spacecraft, allowing the azimuthal angle for a
given scatter to be determined to within
$\Delta\phi=13\deg$\,r.m.s. This angular uncertainty will decrease
potential modulations by a factor of 0.95, which is included in our
calculated modulation factor.

\begin{figure}
\centerline{\includegraphics[width=0.75\textwidth]{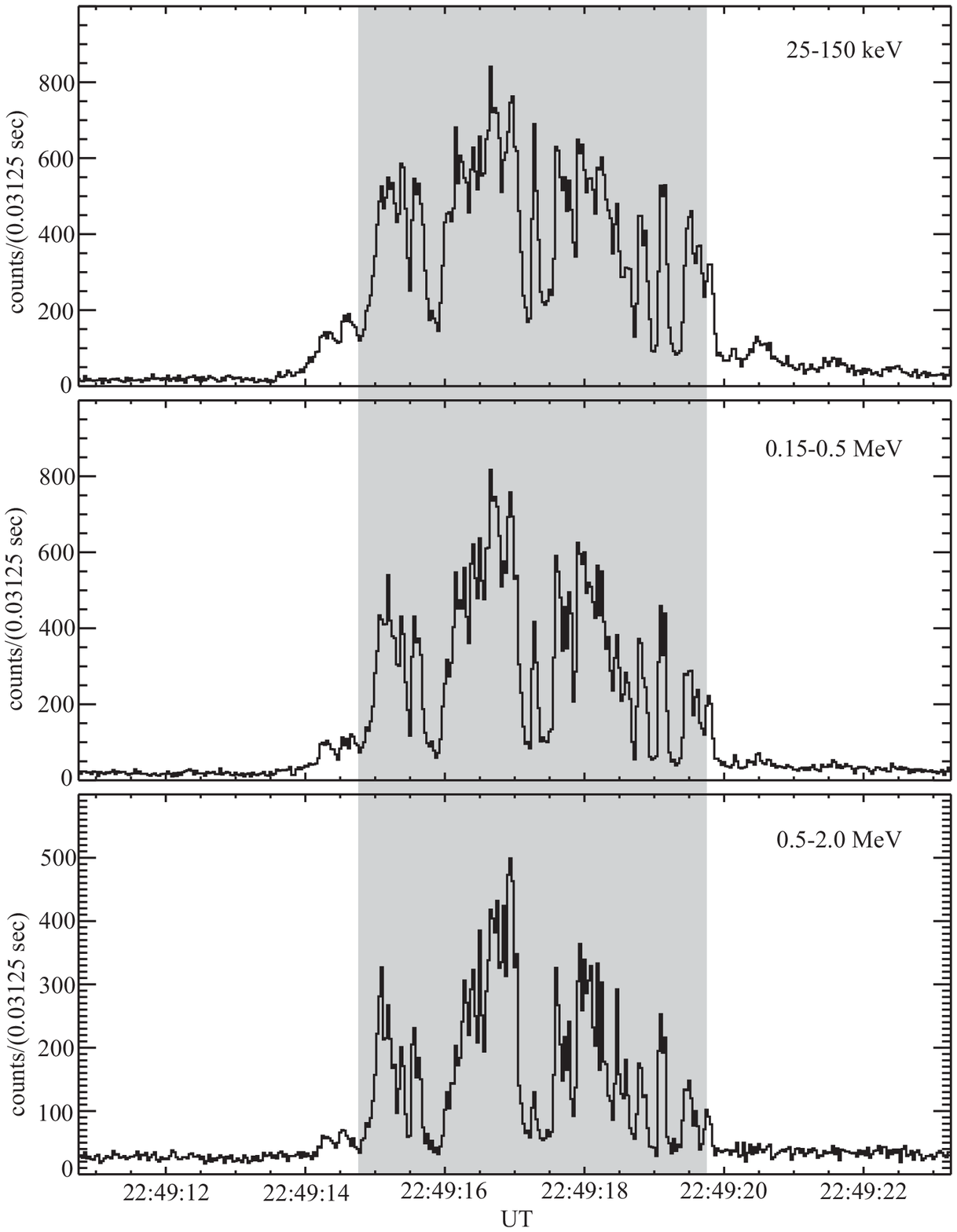}}
\caption{\label{figone} \rhessi\ light curves (in total measured
counts) in three energy bins for GRB21206. The IPN
localized\citep{hur02b} this GRB to 18\deg\ off solar, which precluded
optical afterglow searches; however, the brightness, duration, and
proximity to the \rhessi\ rotation axis made it an ideal candidate to
search for polarization. The shaded region shows our 5\,s integration
time for the polarization analysis.}
\end{figure}

\begin{figure}
\centerline{\includegraphics[width=0.75\textwidth]{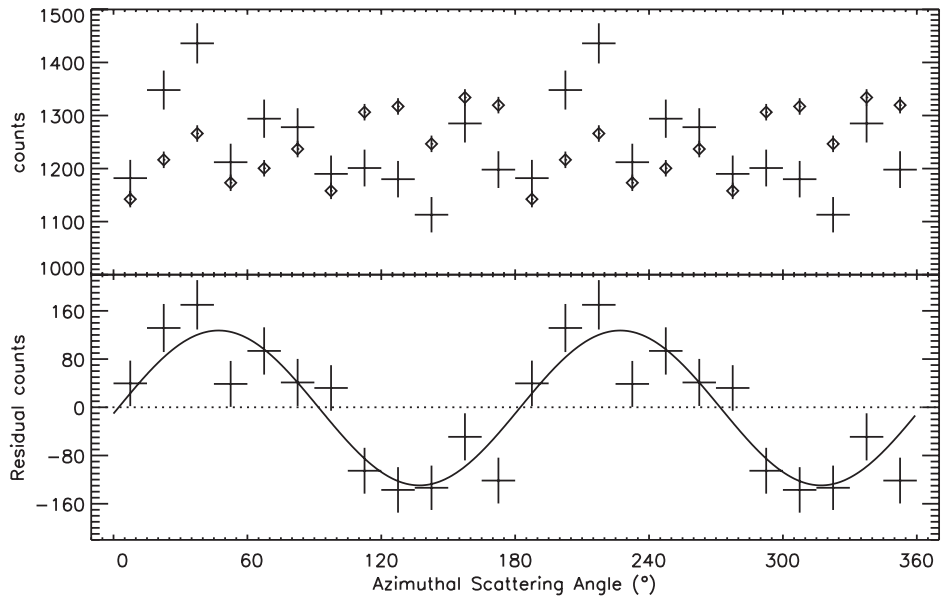}}
\caption{\label{figtwo} The azimuthal scatter distribution
for the \rhessi\ data, corrected for spacecraft rotation. Counts were
binned in 15\deg\ angular bins between 0\deg--180\deg, and plotted
here twice for clarity. The top plot shows the raw measured
distribution (crosses), as well as the simulated distribution for an
unpolarized source (diamonds) as modelled with a Monte Carlo code,
given the time-dependent incident flux. The bottom plot shows the
\rhessi\ data with the simulated distribution subtracted. This
residual is inconsistent with an unpolarized source (dashed line) at a
confidence level $>5.7\sigma$. The solid line is the best-fit
modulation curve, corresponding to a linear polarization of
$(80\pm20)$\%.}
\end{figure}

Prompt $\gamma$-ray emission from GRB021206 was detected with
\rhessi\ on 6 December 2002 at 22:49 UT (Fig.~\ref{figone}). This GRB was
also observed\citep{hur02a} with the Interplanetary Network (IPN),
which reported a 25--100\,keV fluence of
$1.6\times10^{-4}$\,erg\,cm$^{-2}$, and a peak flux of
$2.9\times10^{-5}$\,erg\,cm$^{-2}$\,s$^{-1}$, making this an extremely
bright GRB. The IPN localized\citep{hur02b} GRB021206 to a
57\,square-arcminute error box located 18\deg\ from the Sun. For
\rhessi\, we analysed photons in the energy range 0.15--2.0\,MeV that
scattered between two, and only two, detectors for the 5\,s
integration period shown in Fig.~\ref{figone}. Scattered photons
constitute roughly 10\% of the total 0.15--2.0\,MeV light-curve
events. Counts were binned by the centre-to-centre azimuthal angle
between the two detectors around the \rhessi\ roll axis, corrected for
the rotation of the spacecraft at the time of the photon event. This
azimuthal distribution is plotted in Fig.~\ref{figtwo}. The top panel
shows the raw data, as well as the expected variation for an
unpolarized GRB due only to the light-curve variability. The bottom
panel shows the residual of the data once this unpolarized
distribution is subtracted. The residual shows a large modulation,
which we interpret as a linear polarization of
$\Pi_{m}=(80\pm20)$\%. This observation is the first astrophysical
polarization measurement at $\gamma$-ray energies. Note that the
uncertainty on this polarization amplitude reflects in large part our
uncertainty in the modulation factor. The fact that we have measured a
polarization somewhere in this range has a significance of $<10^{-8}$
(or a confidence $>5.7\sigma$).

Linear polarization is generally considered a clear indication of
synchrotron emission. For electrons with an energy spectrum
characterized by a power-law distribution with spectral index $p$,
synchrotron photons are emitted with a linear
polarization\citep{rybicki79} of $\Pi=(p+1)/(p+7/3)$. For shock
acceleration\citep{bla87}, typical values of $p=$2--3 correspond to
linear polarizations of 70--75\%. For unresolved sources,
polarizations from many directions generally add together to produce
net polarizations that are a fraction of this maximum value. While the
source is unresolved in our observation, if the source electrons are
moving with a bulk Doppler factor of $\Gamma$, we are only viewing the
source over a solid angle $\Omega_{\gamma}\approx1/\Gamma^{2}$. Thus,
for typical values of $\Gamma\geq300$ that have been implied from GRB
afterglow observations, we are effectively resolving a source region
of solid angle $\Omega_{\gamma}\approx10^{-5}$\,sr. Our measurement of
this high polarization is consistent with synchrotron origin for the
initial GRB from a region of nearly uniform magnetic field. For this
emission process to be radiatively efficient as is implied by many
afterglow observations\citep{fra00,fre01}, the magnetic energy
densities must be close to equipartition (comparable to the kinetic
energy densities of the fireball)\citep{der01,gue01}.

This polarization from the prompt $\gamma$-ray emission is
significantly higher than optical polarizations of 1--3\% typically
measured from afterglows\citep{cov99,wij99,rol00}, as well as the
optical polarization of 10\% recently reported\citep{ber03} from the
afterglow of GRB020405. The afterglow emission implies a strong
magnetic field behind the shocks\citep{wax97}, although the field
energy density can be well below equipartition\citep{gal99}. However,
the implied magnetic field strengths are too large to have been due to
a progenitor field being dragged along by the expanding
fireball\citep{med99}, or compression of the ISM magnetic fields in
external shocks\citep{sar96}.  Therefore, the magnetic fields
responsible for the afterglow synchrotron emission are probably
turbulent fields that have built up behind the
shocks\citep{med99,gru99}, which is consistent with the relatively
small optical afterglow polarizations. This locally generated field
invoked for the afterglow has influenced many researchers to consider
the magnetic field responsible for the prompt burst of $\gamma$-rays
as a turbulent, fireball-induced field as well. In the standard
internal shock model\citep{ree94}, the prompt $\gamma$-rays are
produced by synchrotron emission of electrons accelerated to
relativistic energies by shock acceleration, requiring magnetic energy
densities near equipartition in the progenitor environment. However,
late-time turbulent fields have no direct implications on whether the
fields responsible for the prompt $\gamma$-ray emission are
predominantly turbulent or organized\citep{med99,spr01}.  Our
observation conclusively shows that the engine driving the GRB has a
strong, large-scale magnetic field.

Another potential source of polarization would be as follows: if
unpolarized $\gamma$-rays are initially beamed into a small-angle jet,
and then scatter at an angle $\theta$ into our line of sight. The
polarization induced by this Compton scattering\citep{rybicki79} is
$\Pi=(1-\cos^{2}\theta)/(1+\cos^{2}\theta)$ for photon energies below
$\sim$0.1\,MeV, decreasing as the photon energy approaches and exceeds
the electron rest mass, 0.511\,MeV. For our energy band, we estimate a
maximum Compton- induced polarization of 70\% for scatter angles near
90\deg. However, this process would be inefficient, requiring a much
larger $\gamma$-ray energy budget than the synchrotron case. First,
the distribution of scattered $\gamma$-rays would be nearly isotropic,
requiring an energy budget $4\pi/\Omega_{j}$ larger than if the
observed $\gamma$-rays originated from a collimated jet of opening
solid angle $\Omega_{j}$. Second, to maintain such a high polarization
as we observed, the $\gamma$-rays must undergo only a single scatter
into our direction because secondary scatters will erase the induced
polarization from the initial scatter. This condition requires that
the scatter medium be optically thin, $\tau<<1$, and that the
luminosity of the unseen initial $\gamma$-ray beam be larger than the
observed scattered $\gamma$-ray emission by a factor, $1/\tau$.
Because Compton-scattering-induced polarization requires a total
$\gamma$-ray luminosity several orders of magnitude larger than that
implied by synchrotron emission, and elaborate source geometries, the
synchrotron origin for the polarization is preferred.

We suggest that our observation is evidence that the magnetic fields
are actually powering the GRB explosion itself. It has been argued
that a ``passive'' magnetic field--that is, a field dragged from the
surface of central object with a magnetic dipole moment, but not
driving the GRB--could not be strong enough to produce the prompt
$\gamma$-ray emission without an additional locally generated
turbulent magnetic field\citep{spr01}. If this conclusion holds, our
observation is consistent with models of a magnetically driven GRB
fireball, where the driving magnetic field was generated by extracting
the rotational energy of an accretion disk around a central compact
object through differential rotation\citep{tho94,mes97}, by directly
extracting the spin energy of a black hole threaded by magnetic field
lines\citep{bla77}, or by extracting the spin energy of a highly
magnetized neutron star\citep{uso92}. Alternatively, our observation
of a large-scale magnetic field could support models of dynamos in the
post-shock flows\citep{gru01} if the shocks can be shown to be
unstable on large size scales and on timescales comparable to the
prompt $\gamma$-ray emission.

\noindent{\bf Methods} 

The top panel of Fig.~\ref{figtwo} shows the azimuthal
scatter distribution of 0.15--2.0\,MeV events for the 5\,s integration
period shown in Fig.~\ref{figone}, corrected for the rotation of the
spacecraft at the time of each photon event. This distribution is the
sum of three components: the GRB scatter event rate which averages
$820\pm8$ counts per bin, the chance coincidence rate which averages
$374\pm6$ counts per bin, and the background scatter event rate of
$49\pm2$ counts per bin. These rates were determined and confirmed
using event rates before and after the burst, combined with studies of
the readout times of multiple detectors during scattered events, and
were verified independently using our Monte Carlo simulations.
Although the rotation will average out systematic variations in the
scatter angle distribution, we still have to correct for the complex
time profile of the burst itself, which will cause variations for an
unpolarized source owing to the finite number of potential scatter
angles \rhessi\ can measure at any given instant. We modelled this
effect by using the 0.15--2.0\,MeV total count rate in the \rhessi\
instrument (Fig.~\ref{figone}) as the time dependent flux template for
a photon transport Monte Carlo simulation, and using the time-averaged
GRB photon spectrum as measured by \rhessi\ for our input spectrum.
This simulation used the detailed \rhessi\ mass model that has been
developed under CERN's GEANT package, allowing us to model the
instrument response to a GRB at the IPN\citep{hur02b} sky coordinates
for each rotation angle and instantaneous flux, assuming an
unpolarized source. This distribution is also presented in the top
panel of Fig.~\ref{figtwo}. We are looking for a modulation signal
relative to this variation induced by the GRB time profile.

In the bottom panel of Fig.~\ref{figtwo} we show the residual of the
measured distribution once we have subtracted away the simulated
response for an unpolarized GRB, showing our absolute modulation
signal. For an unpolarized source we would expect this distribution to
be flat, which we can rule out at an extremely high confidence level
($\chi^{2}=83.5$, 11\,degrees of freedom, d.f.). When we fit this with
a modulation curve, the fit improves significantly ($\chi^{2}=16.9$,
9\,d.f.), with an amplitude of $128\pm16$ counts per bin.
Statistically, this is a reasonable fit to the data (the probability
of $\chi^2>16.9$ is 5\%), but could be improved with even more
detailed Monte Carlo simulations including timedependent spectral
variability. Using the count rates given above and the simulated
distribution for an unpolarized GRB, we performed further numerical
simulations to determine the probability that an unpolarized GRB could
produce a modulation as large as the one we measure due to random
Poisson counting statistics. We found this probability to be very low,
$<10^{-8}$, which translates to a confidence that we have measured a
polarization at a level $>5.7\sigma$. Finally, we estimated the
modulation factor to be $\mu_{m}=0.19\pm0.04$, using both a separate
photon transport code which fully treats polarization in scattering
and uses a simplified mass model, as well as analytical estimates
based on the GEANT simulation with the full \rhessi\ mass
model. Combining the modulation amplitude, the total source scatter
event rate, and the \rhessi\ modulation factor, we derive a measured
polarization $\Pi_{m}=(80\pm20)$\%.

A number of tests were performed to check that themeasured modulation
is real. First, we verified that the simulated variation induced by
the GRB light curve is accurate by comparing it to an angular
distribution of events that were chance coincidences in two
detectors. These interactions are nearly simultaneous, but separated
by enough time to distinguish them as chance coincidences, not real
scattered photons. This distribution should exhibit the same
variations owing to the GRB light curve, but no polarization.  When we
subtracted the simulated distribution from the chance-coincident
distribution, we found no evidence for a residual modulation. We have
performed a number of independent checks to make sure we do not see
modulations from other sources as well. We have verified that extended
\rhessi\ background observations show no sign of modulations. In
addition, we have done a preliminary analysis of a strong solar
$\gamma$-ray flare observed on 23 July 2002, where we see some
evidence for a modulation, but corresponding to a polarization
$<10$\%. Therefore we feel confident that we have characterized the
systematic effects in \rhessi\ to below the 10\% polarization level.

\vspace*{1em}
\noindent Received 7 February; accepted 12 March 2003

\vspace*{1em}
\noindent{\bf Acknowledgements} We thank D. Smith for help in learning \rhessi\
data analysis and providing simulation support, K. Hurley for IPN data
and references, R. Lin, E. Quataert, J. Arons, C. Matzner and I. Fisk
for discussions, and especially the \rhessi\ team for making all of
their data immediately available to the public at
http://rhessidatacenter.ssl.berkeley.edu.

\vspace*{1em}
\noindent{\bf Correspondence} and requests for materials should be addressed
to S.E.B. (boggs@ssl.berkeley.edu) or W.C. (wcoburn@ssl.berkeley.edu)

\bibliographystyle{unsrt}

\end{document}